\documentclass[preprint,aps]{revtex4}
\usepackage{epsfig}
\usepackage{graphics}
\usepackage{graphicx}

\newcommand{\ybar}{\bar{\psi}}
\newcommand{\n}{\nu}
\newcommand{\y}{\psi}
\newcommand{\half}{{\scriptsize\frac{1}{2}}}
\newcommand{\dslash}{/\!\!\!\partial}
\newcommand{\Ra}{\rightarrow}

\begin{document}

\title{A POSSIBLE CONNECTION BETWEEN \\
MASSIVE FERMIONS AND DARK ENERGY}

\author{T.\ Goldman}\email{ tgoldman@lanl.gov}
\affiliation{Theoretical Division, MS-B283, 
Los Alamos National Laboratory, Los Alamos, NM 87545 USA}

\author{G. J. STEPHENSON, JR.}\email{gjs@phys.unm.edu} 
\author{P. M. ALSING}\email{ alsingpm@hotmail.com}
\affiliation{Dept. of Physics \& Astronomy, University of New Mexico,
Albuquerque, New Mexico 87131, USA}

\author{B. H. J. MCKELLAR}\email{ bhjm@unimelb.edu.au}
\affiliation{School of Physics, University of Melbourne,
Victoria 3010, Australia}

\begin{flushright}
{LA-UR-09-02203}\\
{arXiv:yymm.nnnn}\\
\end{flushright}

\begin{abstract}
In a dense cloud of massive fermions interacting by exchange of a light scalar field, 
the effective mass of the fermion can become negligibly small. As the cloud expands, 
the effective mass and the total energy density eventually increase with decreasing 
density. In this regime, the pressure-density relation can approximate that required 
for dark energy. We apply this phenomenon to the expansion of the Universe with a 
very light scalar field and infer relations between the parameters available and 
cosmological observations. Majorana neutrinos at a mass that may have been recently 
determined, and fermions such as the Lightest Supersymmetric Particle (LSP) may both 
be consistent with current observations of dark energy.
\end{abstract}

\maketitle

\section{Introduction}\label{sec:intro}
Several years ago it was suggested that neutrinos might interact weakly among themselves 
through the exchange of a very light scalar particle~\cite{MKY,TRE}, with possible consequences 
for the evolution of the Universe and for the propagation of neutrinos from distant events.  
We examined such a system for scalars with astrophysical ranges to explore the possibility 
of neutrino clustering~\cite{SGM} and noted at the time that the neutrino clouds thus formed 
could seed structure formation in the Early Universe. More generally, in such clouds of massive 
fermions interacting by exchange of a light scalar field, the effective mass of the fermion can 
become negligibly small. We found that, as a consequence, when the cloud expands, the effective 
mass and the total energy density must eventually increase with decreasing density. We studied 
this system in 1996~\cite{SGM}, well before the discovery of Dark Energy, in connection with 
experimental problems encountered in the search for the mass of the (electron) neutrino. Those 
anomalies have since disappeared, but provided us with the technology to describe dark energy 
in a well understood dynamical system. 

In the following, we first review our previous work on the theory of massive fermions interacting 
via exchange of a scalar field. This is carried out with scaled variables so the regime of applicability 
is not constrained. We next recall the relation between Dark Energy and equations of state and define 
the $w$ parameter used therein. After this, we apply our model results for $w$ and discuss the 
numerical, analytical and scaling properties relevant to the accuracy of our results. Penultimately, 
we extract a rough mass value from applying those results to describe Dark Energy assuming the 
currently accepted value for its energy density in the epoch corresponding to $z = 1$. Finally, we 
present our conclusions and discuss some open questions. 

\section{Summary of a Theory of Massive Fermions \\ 
Interacting via Light Scalar Field Exchange}

The effective Lagrangian for a Dirac field, $\y$, interacting with a scalar field, $\phi$, is:
\begin{equation}
{\cal L} = \ybar(i\dslash - m_\n^{(0)})\y +\half\left[\phi(\partial^2 -
m_s^2)\phi\right] +g\ybar\y\phi
\end{equation}
which gives as the equations of motion
\begin{eqnarray}
\left[\partial^2 + m_s^2\right]\phi & = & g\ybar\y \label{eq:PHI}\\
\left[i\dslash - m_\n^{(0)}\right]\y & = & -g \phi \y.\label{eq:PSI}
\end{eqnarray}
As usual, we set $\bar{h} = c = 1$.
We have omitted nonlinear scalar selfcouplings here, even though they are
required to exist by field theoretic selfconsistency,~\cite{MGGS} as they may
consistently be assumed to be sufficiently weak as to be totally irrelevant.  
The parameter $m^{(0)}_f$ is the renormalized vacuum mass that the fermion 
would have in isolation, and takes into account any contributions from all 
other interactions, as well as contributions from the vacuum expectation value 
of the new scalar field, $\phi$.

We look for solutions of these equations in infinite matter which are static and 
translationally invariant. Eq.(\ref{eq:PHI}) then gives
\begin{equation}
\phi = \frac{g}{m_s^2} \ybar\y,
\end{equation}
which, when substituted in Eq.(\ref{eq:PSI}) gives an effective mass for the fermion 
of
\begin{equation}
m^*_f = m^{(0)}_f - \frac{g^2}{m_s^2}\ybar\y.
\end{equation}

These equations are operator equations. We next act with each of these equations 
on a state $|\Omega \rangle$ defined as a filled Fermi sea, with a number density 
$n$ per fermion state, and Fermi momentum $k_F$, related as usual by 
$n~=~{k_{F}}^{3}/(6\pi^{2})$.  The operator $\ybar\y$ acting on this state gives
\begin{equation}
\ybar\y |\Omega \rangle = \frac{\zeta}{(2\pi)^3}\int_{|\vec{k}|< k_F} d^3k
\left|\frac{m^*_f}{\sqrt{(m^*_f)^2 + k^2}}\right| |\Omega \rangle,
\end{equation}
where $\zeta$ is the number of fermion states which contribute --- $\zeta~=~2$ 
for Majorana fermions and $\zeta~=~4$ for Dirac fermions. Thus the effective mass 
is determined by an integral equation
\begin{equation}
m^*_f = m^{(0)}_f - \frac{g^2\zeta}{2\pi^2m_s^2} \int_0^{k_F} k^2\,dk
\frac{m^*_f}{\sqrt{(m^*_f)^2 + k^2}.}  \label{eq:MSTAR}
\end{equation}

To discuss the solutions of this equation, we reduce it to dimensionless
form, dividing by $m^{(0)}_f$, and introducing the parameter
\begin{equation}
K_0~\equiv~\zeta\frac{g^2(m^{(0)}_f)^2}{2\pi^2m_s^2}, 
\end{equation}
and the variables
$y~=~\frac{m^*_f}{m^{(0)}_f}, x~=~\frac{k}{m^{(0)}_f},
x_F~=~\frac{k_F}{m^{(0)}_f}$.  Then Eq.(\ref{eq:MSTAR})  becomes
\begin{eqnarray}
y& = &1 -  y K_0 \int_0^{x_F} \frac{x^2\,dx}{\sqrt{y^2+x^2}} \\
&=& 1 - \frac{y K_0}{2}\left[e_Fx_F - y^2 \ln\left(\frac{e_F +
x_F}{y}\right) \right],\label{eq:YX}
\end{eqnarray}
with $e_F~\equiv~\sqrt{x_F^2 + y^2}$.  This choice of scaled variables
gives all energies (and momenta) in units of the vacuum fermion
mass.  For consistency, we define the dimensionless scalar mass as
$\mu=\frac{m_s}{m^{(0)}_{f}}$ in these same units. One can regard 
Eq.(\ref{eq:YX}) as a non-linear equation for $y$ as a function of either 
$e_F$ or $x_F$.  As a function of $e_F$, $y$ is multiple valued (when 
a solution exists at all), whereas $y$ is a single valued function of $x_F$.

The total energy of the system is a sum of the energy of the fermions, 
$ E_f~=~e_f~m^{(0)}_{f}~\zeta N$, and the energy in the scalar field,
$E_s~=~e_s~m^{(0)}_{f}~\zeta N$, where $N$ is the total number of
neutrinos in each contributing state.  These expressions serve to define 
the per fermion quantities $e_f$ and $e_s$. Also, $E_s~=~{\cal E}_s~V$, 
where ${\cal E}_s~=~\frac{1}{2}m_s^2\phi^2$ is the energy density of the 
(here uniform) scalar field.

One finds that
\begin{eqnarray}
e_f & = & \frac{3}{x_F^3}\int_{0}^{x_F}x^2\, dx \sqrt{x^2 +
y^2}\nonumber\\
& = & \frac{3}{x_F^3}\left\{\frac{e_F x_F^3}{4} + \frac{e_F x_F y^2}{8} -
\frac{y^4}{8}\ln\left(\frac{e_F + x_F}{y}\right)\right\}
\end{eqnarray}
and
\begin{eqnarray}
e_s & = &
\frac{K_0}{2}\frac{3}{x_F^3}y^2\left(\int_{0}^{x_F}\frac{x^2\,
dx}{\sqrt{x^2 + y^2}}\right) ^2\nonumber\\
& = & \frac{1}{2 K_0} \frac{3}{x_F^3} (1-y)^2.
\end{eqnarray}
and the total energy density per fermion is just the sum, 
\begin{equation}
<e> = e_f + e_s.
\end{equation}

Notice that for large values of $x_F$,
\begin{eqnarray}
y & \Ra & \frac{2}{K_0 x_F^2}\nonumber\\
e_f & \Ra & \frac{3 e_F}{4}\nonumber\\
& \Ra & \frac{3 x_F}{4}\nonumber\\
e_s & \Ra & \frac{3}{2K_0} \frac{1}{x_F^3}.
\end{eqnarray}

It is also useful to note that, for small $x_F$,
\begin{eqnarray}
y & \Ra & 1 - \frac{K_0 x_F^3}{3}\nonumber\\
e_f  & \Ra &  1 + \frac{3 x_F^2}{10}\nonumber\\
e_s  & \Ra &  \frac{K_0}{2}\frac{x_F^3}{3}.\label{eq:LEL}
\end{eqnarray}

\begin{figure}
\includegraphics[scale=0.65]{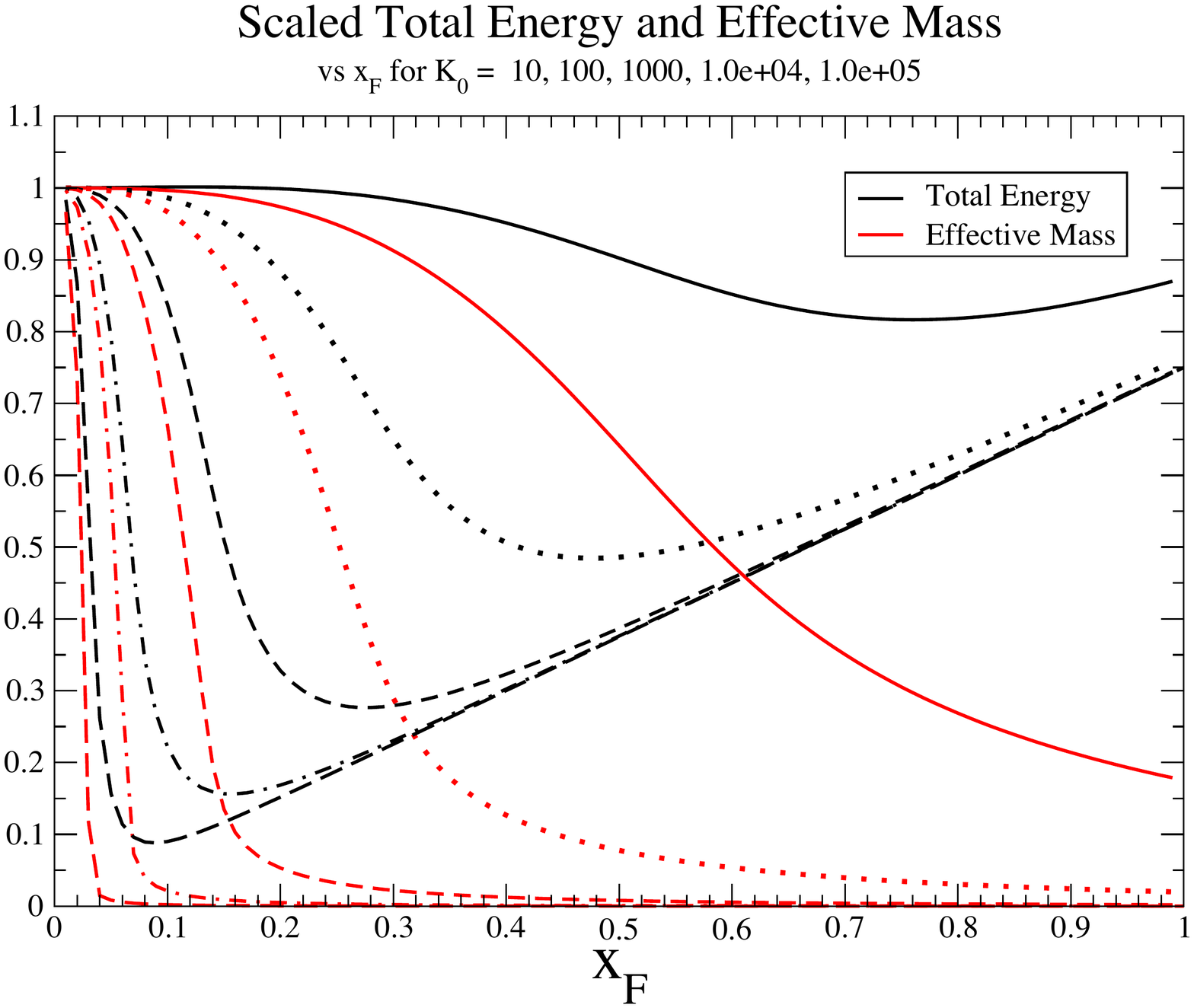}
\caption{Total energy density per fermion $e$ and effective mass $y$
vs. $x_F$ for several values of $K_0$.}
\label{fig:eyvsx}
\end{figure}

For the fermion system to be bound, the minimum of
$e~=~e_f~+~e_s$
as a function of density (or $x_F$) must be less than 1, its value
in the zero density limit. Fig.(\ref{fig:eyvsx}) shows the variation of 
$e$ and $y$ as a function of $x_F$ for several values of $K_0$. 
Note that for sufficiently large $K_0$, there is a minimum relative 
to both the large and small $x_F$ regimes, that is, relative to regions 
of both large and small fermion density. 

\begin{figure}[b]
\includegraphics[scale=0.65]{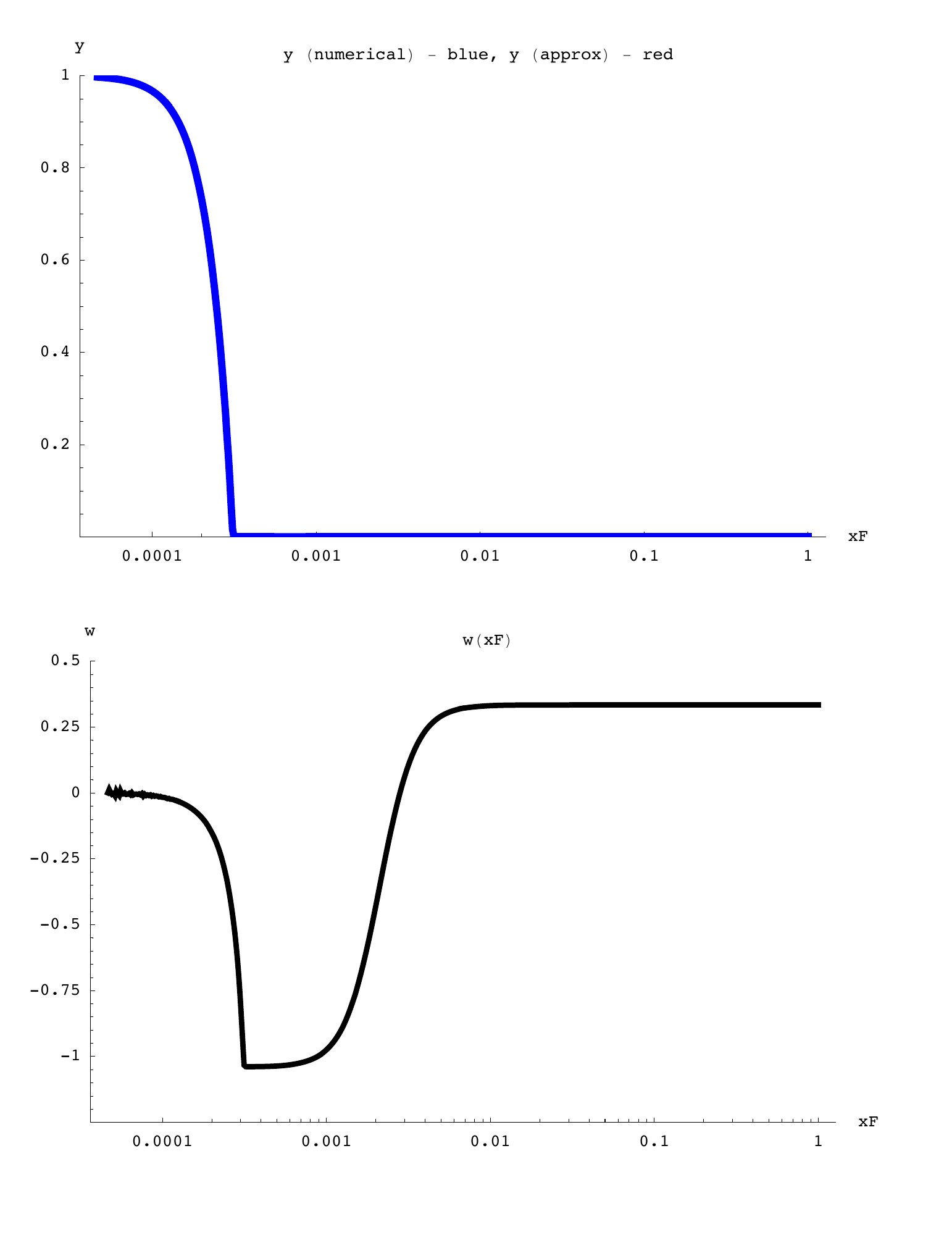}
\caption{$y$ and $w$ vs. log($x_F$) for $K_0 = 10^{11}$.}
\label{fig:ywvsxk11}
\end{figure}

Thinking of this in terms of expansion of the Universe, early times 
correspond to the large $x_F$ region on the right and late times, 
including presumably the present, are to the left. Thus we see that 
at some intermediate period, the energy density passes through a 
minimum and as the system approaches the present, it passes through 
a regime in which the energy density is increasing as the number 
density decreases -- characteristic of a regime of negative pressure. 
Fig.(\ref{fig:ywvsxk11}) gives an advance peek at the value of the equation 
of state parameter, $w$, that we derive from the dependence of $e$ vs. 
$x_F$. We will demonstrate later how we do this numerically, but it should 
be noted that there are still some numerical difficulties evidenced by the 
"hash" in the low $x_F$ limit where we know that $w \Ra 0$ as the cold, 
now non-interacting fermion "dust" turns effectively into new "dark matter". 

\section{Einstein, FRW and Equations of State}

In a Friedmann-LeMa\^{i}tre-Robertson-Walker Universe, Einstein's 
equations produce the second order time derivatiive equation of motion 
that relates the expansion (size) scale parameter, $a$, to Newton's constant, 
$G$, the matter density, $\rho$, pressure, $P$, and a cosmological constant, 
$\Lambda$: 
\begin{equation}
\frac{\ddot{a}}{a} = -\frac{4\pi G}{3} (\rho + 3P) + \frac{\Lambda}{3} 
\end{equation}
Therefore, it is necessary to know the relevant equation of state (EoS) before 
the time development of the scale factor can be determined. For dust, which 
has no pressure, $P = 0$, while for a relativistic gas, $P = \frac{1}{3}\rho$. 
Note that acceleration of the expansion parameter occurs, even in the absence 
of a cosmological constant, when $P < -\frac{1}{3}\rho$. Finally, for a spatially 
and temporally homogenous scalar field, $P < -\rho$, which is more than enough 
to produce acceleration of the expansion. More generally, we can parametrize 
equations of state in this regime by a constant, $w$, as 
\begin{equation}
P = w \rho \label{eq:EoS}
\end{equation}

There are a great number of models for this Dark Energy phenomenon. They go 
by names such as "quintessence" for $ -1 < w < 0$, or "phantom energy" for 
$w < -1$. Among others, this issue has been addressed by Fardon, Nelson and 
Weiner~\cite{FNW}, Peccei~\cite{roberto}, Barshay and Kreyerhoff~\cite{BKmpla}, 
Baushev~\cite{BarX} and Mukhopadhyay, Ray and Choudhoury~\cite{MRC}. 

\subsection{Our EoS}

For the system under consideration, the total (matter plus field) energy is given 
by the product of the total energy density per fermion $<e>$ and the total number 
of fermions, which in turn is determined by the number density $n$ times the volume, 
$V$:  
\begin{equation}
U = \rho V = m^{(0)}_f <e> n V,
\end{equation} 
where we recall that
\begin{equation}
n  =  \zeta \frac{(m^{(0)}_f)^3}{6 \pi^2} x_F^3 
\end{equation}
is the fermion number density.  The pressure is defined by
\begin{equation}
P = -\frac{\partial U}{\partial V},
\end{equation}
where $U$ is the internal energy given above. Since $n V$ is a constant,
\begin{eqnarray}
P & = & - m^{(0)}_f n V \frac{\partial <e>}{\partial V}\nonumber \\
  & = & - m^{(0)}_f n V \frac{\partial n}{\partial V} \frac{\partial <e>}{\partial n}  \nonumber \\
  & = & m^{(0)}_f n^2 \frac{\partial <e>}{\partial n}   \nonumber \\
  & = & \frac{\rho}{3}\frac{x_F}{ <e>} \frac{\partial <e>}{\partial x_F}   \nonumber \\
  & = & \frac{\rho}{3} \frac{\partial ln(<e>)}{\partial ln(x_F)}
\end{eqnarray}

From this and Eq.(\ref{eq:EoS}), we can identify 
\begin{equation}
w = \frac{1}{3}\frac{\partial ln(<e>)}{\partial ln(x_F)}. \label{eq:w}
\end{equation}
This is our central model result. 

\section{General Character of Model Results}

\begin{figure}
\includegraphics[scale=0.65]{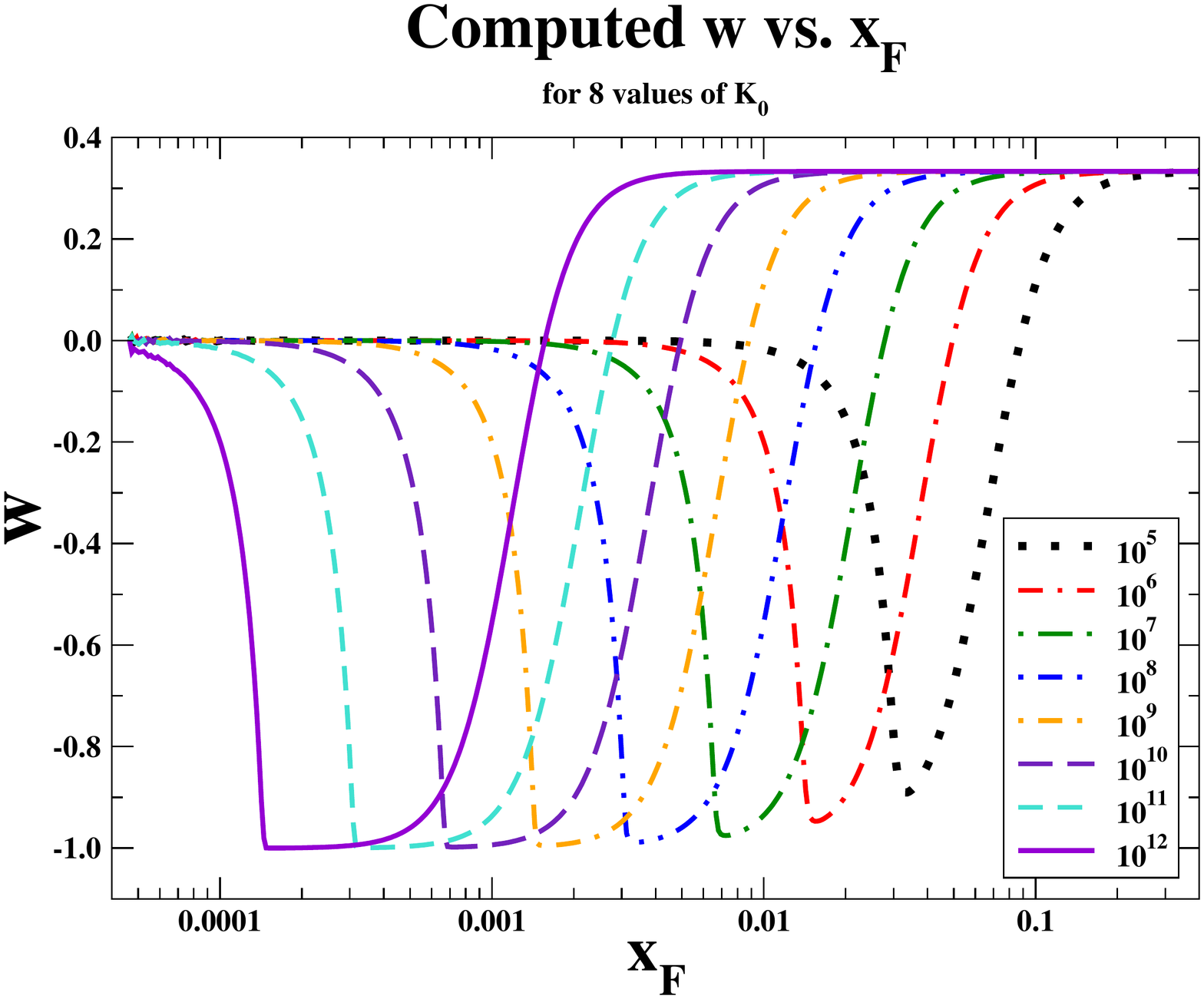}
\caption{$w$ vs. log($x_F$) for 8 values of $K_0$.}
\label{fig:wvslogxf}
\end{figure}

\begin{figure}
\includegraphics[scale=0.65]{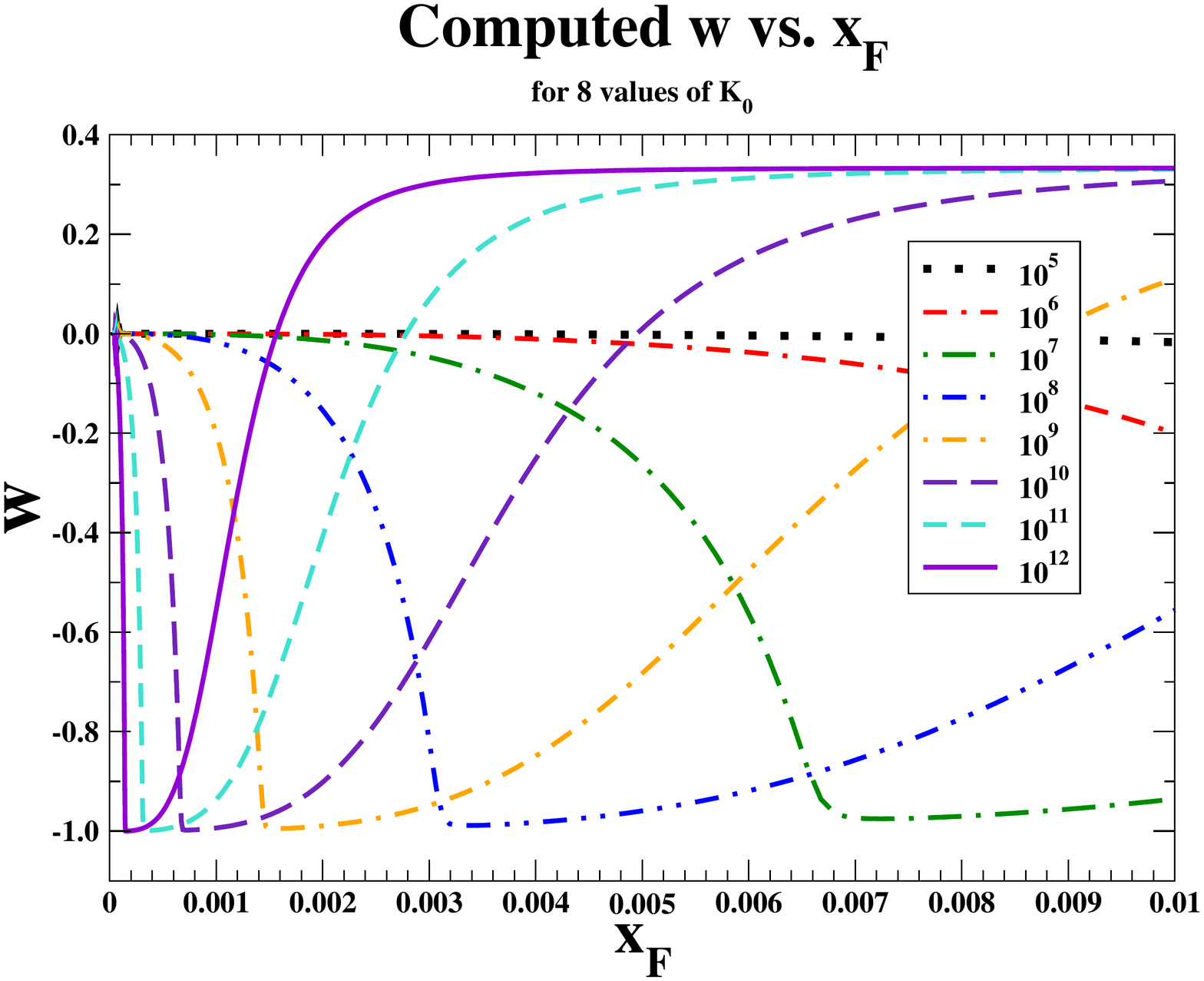}
\caption{$w$ vs. $x_F$ for 8 values of $K_0$.}
\label{fig:wvslinxf}
\end{figure}

In Fig.(\ref{fig:wvslogxf}), we shows the value of $w$ as computed numerically 
from Eq.(\ref{eq:w}) for 8 values of $K_{0}$ on a log scale for $x_F < 0.4$ and 
in FIg.(\ref{fig:wvslinxf}) for $x_F < 0.1$ on a linear scale. Note that it approaches 
close to $-1$ as the density decreases (as the Universe expands and the scale 
factor $a$ increases from right to left) and then departs sharply towards zero, 
as also indicated earlier in Fig.(\ref{fig:ywvsxk11}). 

At large $x_F$, it is clear that $w$ approaches $+1/3$ as it should for a relativistic 
gas of fermions. It is perhaps less clear, due to numerical fluctuations, that the value 
goes to zero at zero density. This can be checked by considering the small $x_F$ 
expansions of the energy densities shown at Eq.(\ref{eq:LEL}). 

\begin{figure}
\includegraphics[scale=0.65]{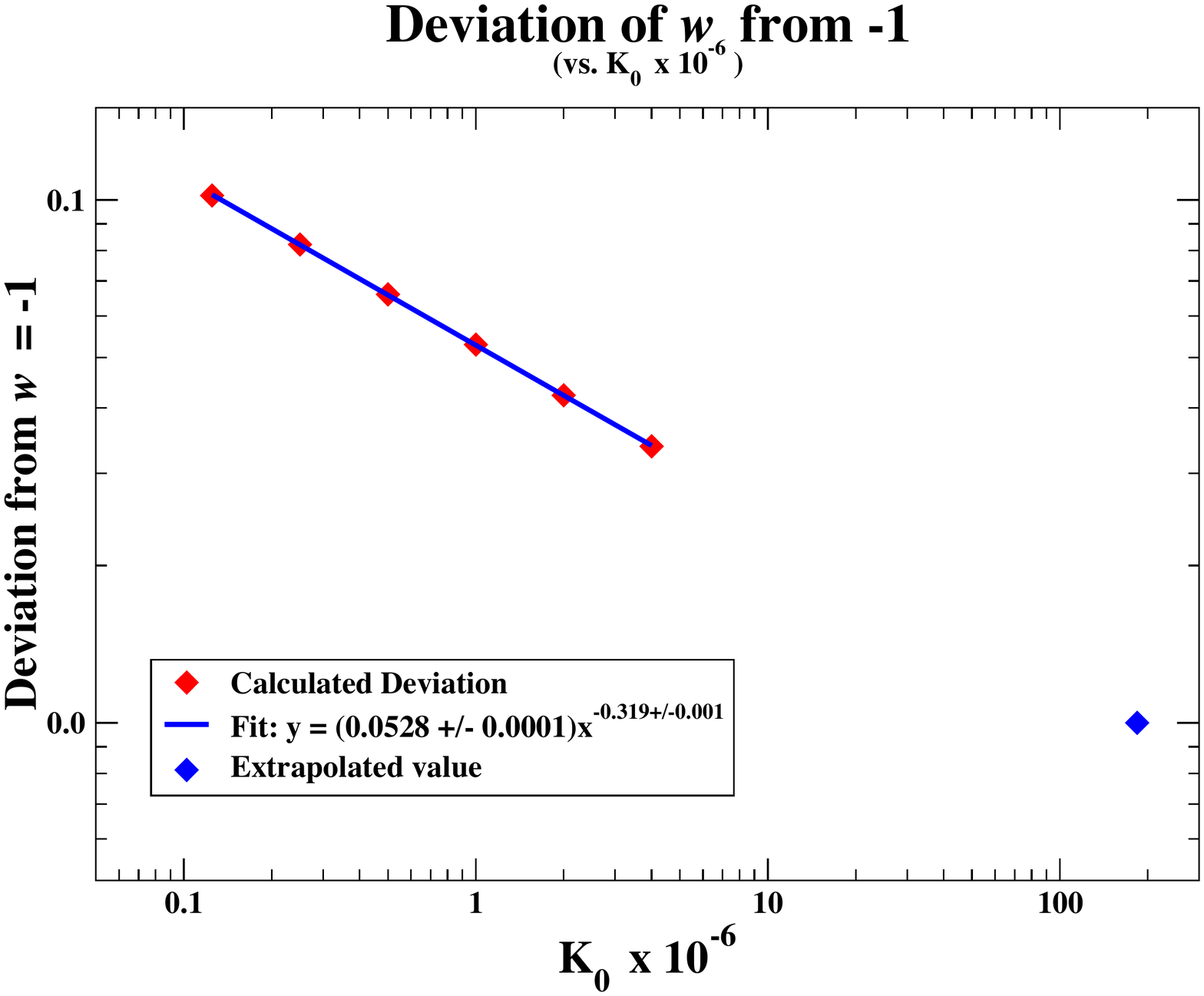}
\caption{Deviation of minimum value of $w$ from $-1$ vs. $K_0$.}
\label{fig:wmin1}
\end{figure}

\begin{figure}
\includegraphics[scale=0.65]{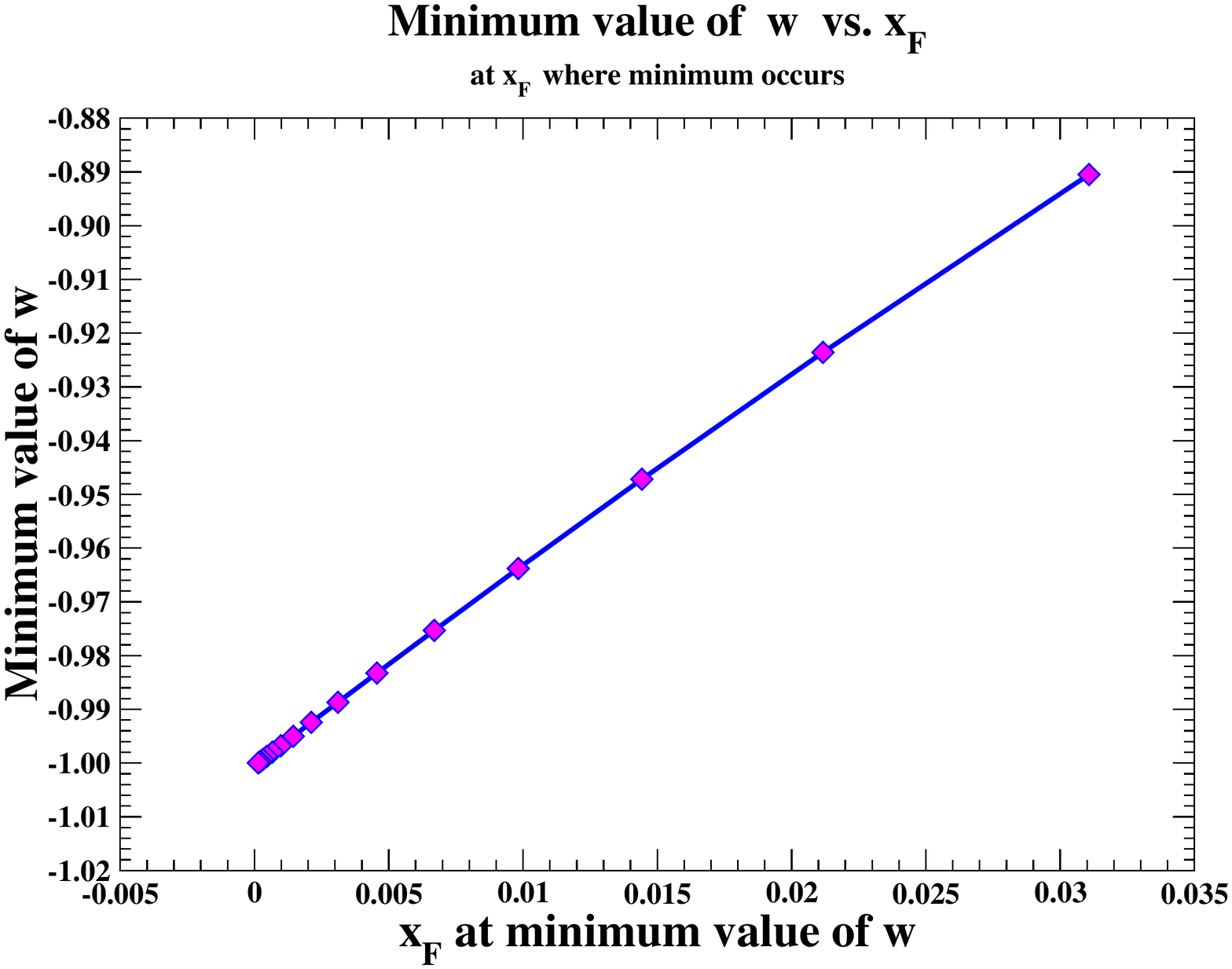}
\caption{Minimum value of $w$ vs. $x_F$ at which minimum occurs, 
for several values of $K_0$.}
\label{fig:wmin2}
\end{figure}

\begin{figure}
\includegraphics[scale=0.65]{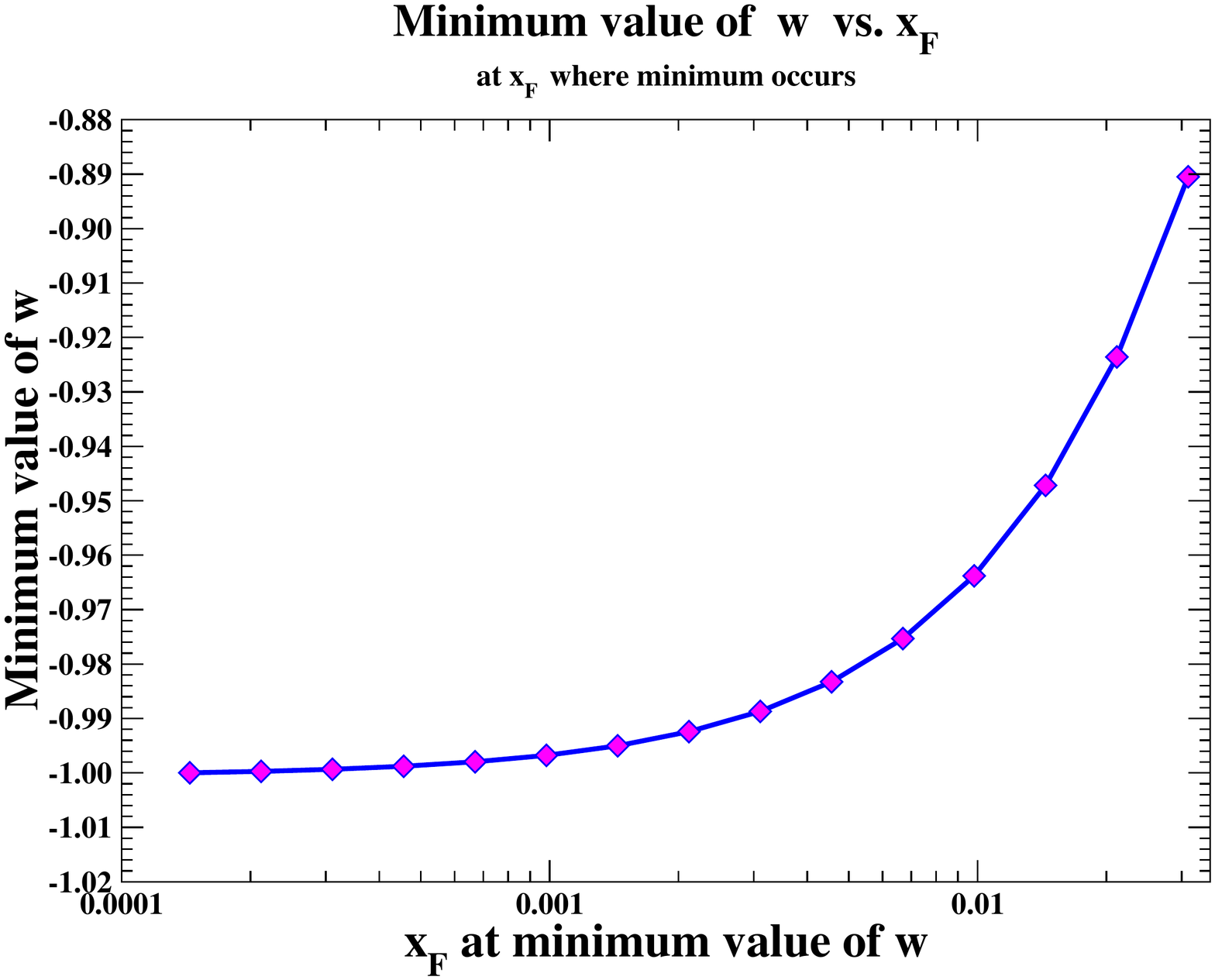}
\caption{Minimum value of $w$ vs. logarithm of $x_F$ at which minimum occurs, 
for several values of $K_0$.}
\label{fig:wmin3}
\end{figure}

Finally, we performed a number of numerical scaling checks to examine whether $w$ 
can fall below $-1$. The are shown in Figs.(\ref{fig:wmin1},\ref{fig:wmin2},\ref{fig:wmin3}).

We are continuing our efforts to ensure numerical stability primarily by seeking 
analytic formulae and approximations, especially to avoid taking derivatives numerically. 
We will report on these improvements elsewhere, but suffice it to say that we have 
found no contradiction to the results obtained here by purely numerical means, and 
confirmed that the difficulties encountered at very small $x_F$ are indeed due to 
numerical noise. 

\section{Numerical Values}

All of the above is carried out with scaled variables. However, was shown in 
Ref.[3], there are only a few, weak constraints on the actual parameter 
values. Very large values of $K_0$ are possible even for very small values of 
$g^2$ if the range of the scalar is very large, corresponding to very small values 
of $m_s$. Even if long-ranged, such weak interactions between fermions, especially 
neutrinos or those outside of the Standard Model altogether (such as the LSP) are 
exceptionally difficult to constrain by any laboratory experiments. 

The energy density of Dark Energy is quoted~\cite{density}  as $(3.20  \pm 0.4 ) 
\times 10^{-47} \mbox{Gev}^4$. In more manageable units, this is given as 
$(2.3 meV)^4$. What does this imply for the allowed value of $m_0$? If we set 
this energy density equal to that of this system at $w = -1$, 
\begin{equation}
\rho_{\Lambda}  =  \frac{(m^{(0)}_f)^4 <e> x_F^3}{6 \pi^2} 
\end{equation}
then solving for $m^{(0)}_f$ gives 
\begin{equation}
m^{(0)}_f =  \{\frac{6 \pi^2 \rho_{\Lambda}}{<e> x_F^3}\}^{1/4}
\end{equation}
If we further suppose that this occurs at cosmological $z \sim 1$, then the 
range of the scalar field must be comparable to the size of the Universe at 
that time. That is, 
\begin{equation}
m_s \sim 7 \times 10^9 \, {\rm light years} \sim 3 \times 10^{-30} \, {\rm meV}.
\end{equation}
For the relatively modest value of $K_0 \sim 8 \times 10^{6}$, this implies 
that $m^{(0)}_f \sim 300 {\rm meV}$ and $g^2/(4\pi) \sim 6 \times 10^{-58}$. 
These values emphasize the virtual impossibility of constraining this physics by 
means of laboratory experiments. 

We note with interest that following the curve for $y$ from $z = 1$ to "now", 
i.e., $z = 0$, tells us that the effective mass of this fermion would now be measured 
to be approximately  $(7/8)m^{(0)}_f$, a value tantalizingly close to the Majorana 
neutrino mass that Prof. Klapdor has reported from his experiments~\cite{Klap}. 

\begin{figure}
\includegraphics[scale=0.65]{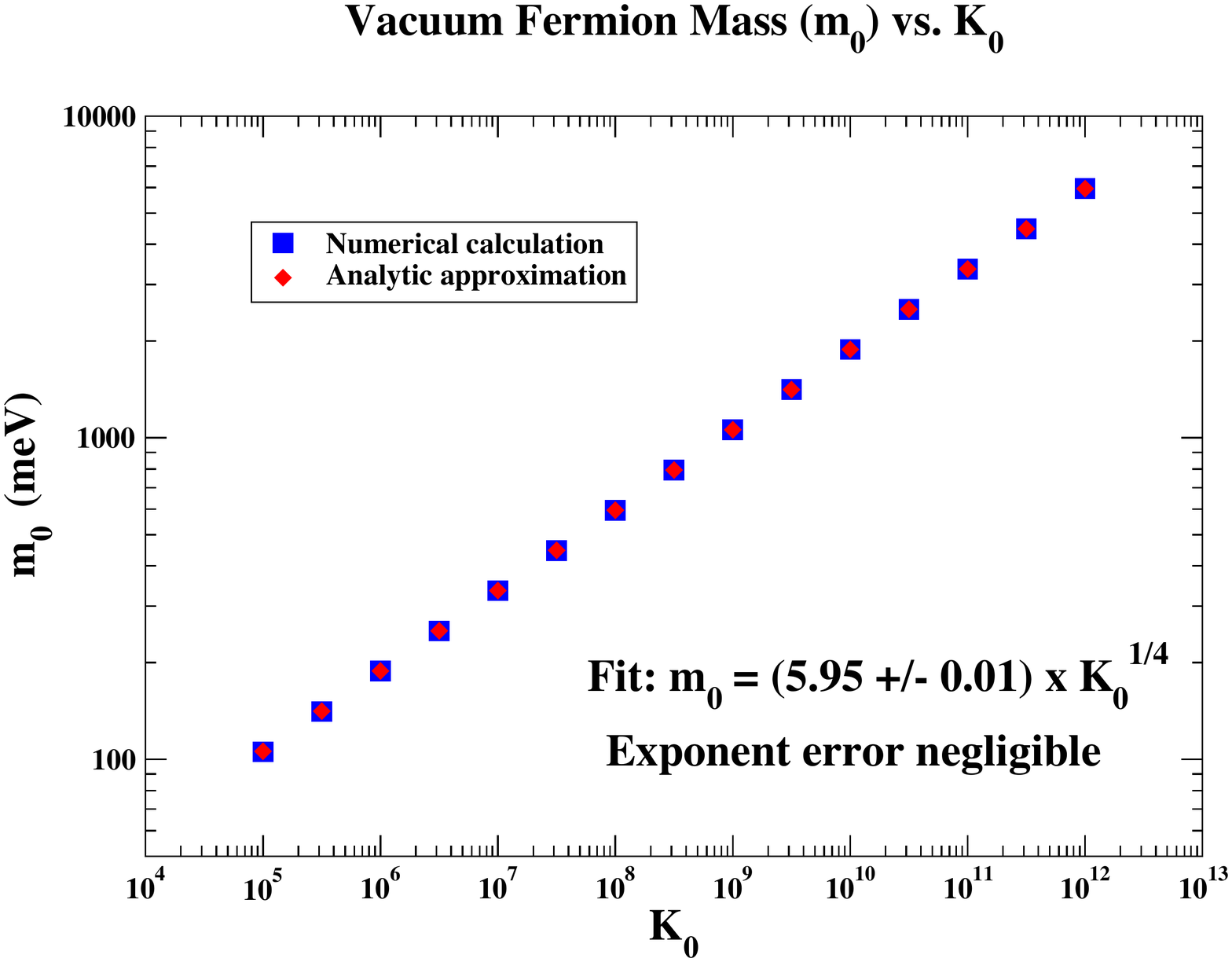}
\caption{Scaling variation of the vacuum fermion mass, $m^{(0)}_f$, vs. $K_0$.}
\label{fig:scaling}
\end{figure}

Other solutions are possible, and Fig.(\ref{fig:scaling}) shows how our results scale 
very accurately (for sufficiently large $K_0$) with  the 4th root of $K_0$, both 
numerically and under one of our analytic approximations to the region where $w$ 
is a minimum. In particular, if one has the LSP at $\sim 1$ TeV in mind, $K_0$ 
becomes very large, $\sim 10^{57}$ and $g^2/(4\pi)$ also increases, $\sim 10^{-32}$, 
but these values are not ruled out by anything known. 

\section{Conclusions and Questions}

We have displayed an explicit and calculable dynamical mechanism that describes 
Dark Energy and connects it to what in the current epoch becomes a kind of dark 
matter. Although Dark Matter is usually thought of as existing at early epochs in the 
life of the Universe, the system described here turns (hot) relativistic fermions, that 
interact weakly with a very light scalar field, into Dark Energy which lasts for a limited 
time during the expansion of the Universe which then morphs into new cold dark matter 
components. Thus, neutrinos can contribute to both. Nor need there be only one time 
scale or one species for which this applies. (See, e.g., Ref.[12].) If there are 
many sterile fermions with sufficiently long decay lifetimes, the acceleration/deceleration 
history of the Universe could be much more complicated that presently envisioned. 

We may also ask generally why $w > -1$, but it is fairly clear in this model: A scalar 
field strength uniform in space and time produces~\cite{CHT} exactly $w = -1$, but 
here the source for the scalar field is the density of massive fermions. Relativistically, 
their strength for producing scalar field is severely reduced at high momentum (in the 
rest frame of the Universe) but as they slow, the scalar field strength grows nonlinearly 
until, due to the expansion of the Universe, the fermions separate so much (greater than 
the Yukawa range for exchange of the scalar field) that they cannot act collectively and 
the scalar field strength declines again. 

Finally, we note that our model has definitive if difficult tests, as it predicts specific 
variations of $w$, slowly approaching $-1$ from above as $z$ decreases through 1, 
and rising rapidly towards zero as $z$ approaches the present. We hope this encourages 
observationalists in their efforts to discern variation of $w$ with $z$. 

\section*{Acknowledgments}|
This work was carried out in part under the auspices of the National Nuclear Security
Administration of the U.S. Department of Energy at Los Alamos National Laboratory
under Contract No. DE-AC52-06NA25396 and supported in part by the Australian
Research Council.

\end{document}